\documentclass[oribibl]{llncs}
\usepackage{multicol}
\usepackage{amsmath}
\usepackage{amssymb}
\usepackage{graphicx}
\usepackage{color}
\usepackage{hhline}

\newcommand{\ignore}[1]{}

\newcommand{\mf}[1]{\mathfrak{ #1}}
\newcommand{\mbf}[1]{\mathbf{ #1}}
\newcommand{\e}{\mathbf{e}}
\newcommand{\msf}[1]{\mbox{\sf #1}}

\newcommand{\boxtheorem}{  \hfill $\Box$}
\newcommand{\nit}[1]{{\it #1}}

\newcommand{\mc}[1]{\mathcal{ #1}}

\newcommand{\sfd}{{\sf d}}
\newcommand{\sfs}{{\sf s}}

\title{\vspace*{-8mm}Answer-Set Programs  for  Repair Updates and Counterfactual Interventions \vspace{-4mm}}

\author{{\bf Leopoldo Bertossi}\thanks{Millennium Institute for Foundational Research on Data (IMFD, Chile) \& Professor Emeritus Carleton University, Canada. \  bertossi@scs.carleton.ca}\vspace{-2mm}}
\institute{{\bf Skema Business School, Montreal, Canada}\vspace{-5mm}}

\begin{document}

\thispagestyle{empty}
\pagestyle{plain}
\maketitle

\begin{abstract}\vspace{-2mm}
We briefly describe -mainly through very simple examples- different kinds of answer-set programs with annotations that have been proposed for specifying: database repairs and consistent query answering; secrecy view and query evaluation with them; counterfactual interventions for causality in databases; and counterfactual-based explanations in machine learning. \vspace{-3mm}
\end{abstract}

\paragraph{\bf \em Repair Programs and Consistent Query Answering.} \ When a relational database fails to satisfy a given set of integrity constraints (ICs) and one wants to obtain semantically correct query answers, one can specify and simultaneously query virtually repaired, i.e. consistent, versions of the database. The so obtained certain answers are considered to be the consistent answers \cite{pods99,bertossiSynth}. Answer-set programs (ASPs) with annotations can be used for this task \cite{barcelo1,monica}. The motivation for  this approach comes from earlier work where {\em annotated predicate logic}, a form of multi-valued logic extended with non-monotonic negation, that was used for this task \cite{dood}.

\vspace{-2mm}
\begin{example}  \label{ex:kappa2} \ The following is a {\em denial constraint} (DC), i.e. it prohibits combinations or joins of database atoms: \
$\kappa\!:   \neg \exists x\exists y( S(x)\wedge R(x, y)\wedge S(y))$. \
The following database instance $D$ violates $\kappa$.

\vspace{-3mm}
\begin{multicols}{2}
\begin{center}
{\small {\begin{tabular}{l|c|c|} \hline
$R$  & A & B \\\hline
$\iota_1$ & $a_4$ & $a_3$\\
$\iota_2$& $a_2$ & $a_1$\\
$\iota_3$& $a_3$ & $a_3$\\
 \hhline{~--}
\end{tabular} \hspace*{0.5cm}\begin{tabular}{l|c|c|}\hline
$S$  & A  \\\hline
$\iota_4$& $a_4$ \\
$\iota_5$& $a_2$ \\
$\iota_6$& $a_3$ \\ \hhline{~-}
\end{tabular} } }
\end{center}

We use global tuple identifiers (tids) to refer to individual tuples. They appear in predicates' first arguments followed by a semicolon.
\end{multicols}

\vspace{-4mm}Subset-repairs (S-repairs) of $D$ are consistent w.r.t. $\kappa$, and minimally differ from $D$ under set inclusion. They are: \ $D_1=$ $ \{R(\iota_1;a_4,a_3),$ $ R(\iota_2;a_2,a_1),$ $ R(\iota_3;a_3,a_3),$ $ S(\iota_4;a_4), S(\iota_5;a_2)\}$, \  $D_2 = \{ R(\iota_2;a_2,a_1), S(\iota_4;a_4),$ $S(\iota_5;a_2),$ \linebreak $S(\iota_6;a_3)\}$, \ and  $D_3 =$ $\{R(\iota_1;a_4,a_3), R(\iota_2;a_2,a_1), S(\iota_5;a_2),S(\iota_6;a_3)\}$.

The repair program contains the atoms in $D$,  and the rules:

\vspace{-7mm}
\begin{eqnarray*}
S'(t_1;x,\sfd) \vee R'(t_2;x,y,\sfd) \vee S'(t_3;y,\sfd) &\leftarrow& S(t_1;x), R(t_2;x, y), S(t_3;y). \\
S'(t;x,\sfs) &\leftarrow& S(t;x), \ \nit{not} \ S'(t;x,\sfd). \nonumber\\
R'(t;x,y,\sfs) &\leftarrow& R(t;x,y), \nit{not} \ R'(t;x,y,\sfd).\nonumber
\end{eqnarray*}

\vspace{-3mm}
 Here, $t_1, \ldots, t, x, y, \ldots$ are variables, but the annotation $\sfd$  is a constant indicating that the tuple is deleted from $D$. Annotation constant $\sfs$ indicates that the tuple stays in the repair. Here, the first rule captures in its body (i.e. antecedent) a violation of $\kappa$, and the head (i.e. the consequent) offers the alternative tuple deletions that can solve the violation. \ The last two rules specify that repairs keep the original tuples that have not been deleted. Predicates $R'$ and $S'$ are nicknames for $R$ and $S$, with an extra argument for the annotation.

This repair-program has three stable models, with repair $D_1$ corresponding to the model $M_1 = \{R'(\iota_1;a_4,a_3,\sfs), R'(\iota_2;a_2,a_1,\sfs),$ $ R'(\iota_3;a_3,a_3,\sfs), S'(\iota_4;a_4,\sfs),$ $ S'(\iota_5;a_2,\sfs),$ $
S'(\iota_6;a_3,\sfd)\} \cup D$, in the sense that $D_1$ can be read off from $M_1$ by keeping only the tuples annotated with \sfs. \boxtheorem
\end{example}

\vspace{-2mm}
If there are interacting ICs, for which repair actions for one of them may affect satisfaction of another one, it is necessary to use a couple of extra annotations to capture a transition process \cite{monica}. Doing CQA becomes skeptical (or certain) reasoning with the repair program.   A system for CQA based on repair programs that runs on the \nit{DLV} system \cite{leone} is described in \cite{monica}.

In some cases, it is more convenient to specify and compute S-repairs that are also {\em cardinality repairs} (C-repairs), i.e. that differ from the original instance by a {\em minimum number} of tuples \cite{bertossiSynth}. In Example \ref{ex:kappa2}, (only) they can be obtained by  adding to the program the following {\em weak program constraints}  \ (WCs): \ $\sim  R(t,\bar{x}),  R'(t,\bar{x},\sfd)$ \ and \
$\sim  S(t,\bar{x}),  S'(t,\bar{x},\sfd)$. \ Violations, i.e. instantiations of what comes after $\sim$, should not happen. However, they are accepted as long as they are kept to a minimum \cite{leone}.  With these WCs the number of  deleted tuples is minimized.

Repair programs are quite general and can be produced for any set of first-order ICs.  Their complexity and expressive power are just right for the intrinsic complexity of repair-related computational tasks and CQA \cite{monica,bertossiSynth}. Repair programs have been extended for CQA from {\em virtual data integration systems} \cite{bravoJournal}, and to  {\em peer data exchange systems} \cite{tplp17}. Other ASP-based approaches to database repairs and CQA can be found in \cite{tplp,grecoLP,eiterLoc}.

\vspace{-2mm}
\paragraph{\bf \em Virtual Updates for Secrecy Views.} \  Repairs that replace attribute values by null values that behave as in SQL \cite{tkde,tplp} can also be used to hide the contents a  view. Intuitively, a database, for which the contents a view is expected to stay hidden, offers as query answers only the certain (or consistent) answers from the set of {\em secrecy instances} (the repairs) that have the view empty or containing tuples with nulls only. Details about these repair and {\em secrecy} programs can be found in \cite{tkde,kais}.

\vspace{-1mm}
\begin{example} (example \ref{ex:kappa2} cont.) \label{ex:attrib} Consider the same database ${D}$, and $V_{\!\kappa}(x,y): S(x)
\land R(x, y) \land S(y)$ defining the view whose contents we want to hide from users. In order to to do this, we repair w.r.t. the associated DC $\kappa$
 by ``minimally" changing  attribute  values by a constant {\sf NULL}. Since it behaves as in SQL, it cannot be used to
satisfy a join. \ There are two  repairs.

\begin{center}{\small \begin{tabular}{l|c|c|} \hline
$R$  & A & B \\\hline
$\iota_1$& $a_4$ & $a_3$\\
$\iota_2$& $a_2$ & $a_1$\\
$\iota_3$& $a_3$ & $a_3$\\
 \hhline{~--}
\end{tabular} \hspace*{0.2cm}\begin{tabular}{l|c|c|}\hline
$S$  & A  \\\hline
$\iota_4$& $a_4$ \\
$\iota_5$& $a_2$ \\
$\iota_6$& ${\sf NULL}$ \\ \hhline{~-}
\end{tabular}\hspace{15mm}\begin{tabular}{l|c|c|} \hline
$R$  & A & B \\\hline
$\iota_1$ & $a_4$ & ${\sf NULL}$\\
$\iota_2$& $a_2$ & $a_1$\\
$\iota_3$& $a_3$ & ${\sf NULL}$\\
 \hhline{~--}
\end{tabular} \hspace*{0.2cm}\begin{tabular}{l|c|c|}\hline
$S$  & A  \\\hline
$\iota_ 4$& $a_4$ \\
$\iota_5$& $a_2$ \\
$\iota_6$& $a_3$ \\ \hhline{~-}
\end{tabular} }
\end{center}

In each of them {\sf NULL} is preventing the satisfaction of a
join in $V_{\!\kappa}$. In both cases, the set of value changes is minimal under set inclusion.

A program rule that achieves this result is: \
$S'(t_1,{\sf NULL}) \vee R'(t_2,{\sf NULL}, y) \vee R'(t_2,x, {\sf NULL}) \vee S'(t_3,{\sf NULL}) \leftarrow S(t_1,x),R(t_2,x, y),S(t_3,y)$.  \boxtheorem
\end{example}

\vspace{-6mm}
\paragraph{\bf \em Counterfactual Programs for Causality in Databases.} \ In \cite{suciu}, {\em actual causality} \cite{HP05} was applied to define and compute database tuples that are causes for a query to be true. Furthermore, {\em causal responsibility} \cite{CH04} was used to assign numerical scores to causes, to reflect their strength as such. A detailed analysis of causality in databases was carried out in \cite{tocs}, where a useful connection with database repairs was unveiled. As consequence, repair program can be used to specify and compute causes \cite{kais}.

\vspace{-2mm}
  \begin{example} \ (example \ref{ex:kappa2} cont.) \label{`first'} \ With the same database ${D}$,
 the query \linebreak ${\mc{Q}_{\kappa}\!: \ \exists x \exists y ( S(x) \land R(x, y) \land S(y))}$ \ is true in $D$. \
Tuple $t_6$ is a {\em counterfactual cause} for $\mc{Q}_{\!\kappa}$ to be true in ${D}$, in the sense that if $t_6$ is removed from ${D}$,
 $\mc{Q}_{\!\kappa}$ is no longer true. Its  {\em responsibility} is $1$, the highest possible. \
Tuple $t_1$ is an {\em actual cause} since deleting it from $D$ {\em together with} its {\em contingent} tuple $t_3$  makes the query false.
Its responsibility is \  $\frac{1}{1 + |\Gamma|} = \frac{1}{1 + 1} =  \frac{1}{2}$, with $\Gamma$ the smallest set of its contingent tuples.
\ Similarly, $t_3$ and $t_4$ are actual causes, with responsibility  ${\frac{1}{2}}$.

 When $Q_{\!\kappa}$ is true, equivalently the IC $\kappa$ is false, in order to obtain causes, tuples have to be deleted from $D$, for which a repair program for  $\kappa$ can be used. Causes' tids can be retrieved by means of the rules: $\nit{Cause}(t) \leftarrow R'(t,x,y,{\mathsf d})$
and $\nit{Cause}(t) \leftarrow S'(t,x,{\mathsf d})$. \ In order to obtain their contingency tuples, one can use, e.g. $\nit{Cont}(t,t') \leftarrow  R'(t,x,y,{\mathsf d}), R'(t',u,v,{\mathsf d}), t\neq t'$, collecting tids that have to de deleted together with $t$. \ Using set-building and aggregations, one can compute contingency sets and their cardinalities \cite{calimeri08,calimeri09}, and then, also responsibilities. With WCs one can concentrate on minimum cardinality contingency sets \cite{kais}.
 \boxtheorem
\end{example}

\vspace{-4mm}\paragraph{\bf \em Counterfactual Programs for Explainable ML.} \ Consider entity records represented by atomic formulas  $E(t;\bar{e},\mathsf{o})$, with an id $t$, a sequence of feature values $\bar{e}$, and an annotation $\msf{o}$ for ``original record". They are labeled, say with $0$ or $1$, by a black-box or open-box classifier represented by a predicate $\nit{Cl}(t,\ell)$: record with id $t$ received label $\ell$. \ A particular entity $E(\mbf{t};\mbf{\bar{e}};\mathsf{o})$ receives label $1$, i.e. $\nit{Cl}(\mathbf{t},1)$ is true, and we want to explain this by {\em counterfactually intervening} $\mbf{\bar{e}}$, changing feature values, trying to obtain label $0$.  For each feature we do this, but value changes for other features may be necessary (as with actual causality above). A responsibility score can be assigned to each feature value that depends on the additional required changes (similar but much more general than causal responsibility above) \cite{deem,rw21}. These are also called score-based {\em attributive} or {\em contrastive explanations} in explainable AI.

The process of iteratively intervening an entity until it switches label can be specified by means of {\em counterfactual ASPs} \cite{tplp,ijclr21}.  For the gist, we give  some of the rules in such a program. Since an entity may go trough an iteration of feature value changes, we need an annotation, $\mf{\star}$, to indicate it is in transition, with the rules: \ $E(\e;\bar{x};\mf{\star}) \leftarrow E(\e;\bar{x};\msf{o})$ and
$E(\e;\bar{x};\mf{\star}) \leftarrow E(\e;\bar{x};\msf{do})$, where annotation $\msf{do}$ indicates a single counterfactual change.

The main rule, \ $\mbox{\phantom{ooo}}E(t;x_1',x_2, \ldots,x_n;\msf{do}) \vee \cdots \vee E(t;x_1,x_2, \ldots, x_n';\msf{do})  \ \longleftarrow$ $ E(t;\bar{x};\mf{\star}), \nit{Cl}[t;1], \nit{Dom}_1(x_1'), \ldots, \nit{Dom}_n(x_n'),x_1'\neq x_1, \ldots, x_n'\neq x_n,   \nit{choice}(\bar{x};x_1'),$ $\ldots, \nit{choice}(\bar{x};x_n')$, specifies that while the label is not switched,  a single feature value is non-deterministically \cite{zaniolo} replaced by a new one from the feature domain. \ One eventually stops when the label has been switched: \ $E(t;\bar{x};\msf{s}) \ \leftarrow \ E(t;\bar{x};\msf{do}), \nit{Cl}(t,0).$

 \vspace{3mm} \noindent {\bf Acknowledgments: } \  Part of this work was funded by ``ANID - Millennium Science Initiative Program" - Code ICN17002; and ``Centro Nacional de Inteligencia Artificial" CENIA, FB210017, Financiamiento Basal para Centros Científicos y Tecnológicos de Excelencia de ANID, Chile.


\begin{thebibliography}{10}



 \bibitem{pods99} Arenas,~M., Bertossi,~L. and Chomicki,~J. \ Consistent Query Answers in Inconsistent Databases. In Proc. ACM PODS 1999,  pp. 68-79.

\bibitem{dood}
Arenas,~M., Bertossi,~L. and Kifer,~M. \
Applications of Annotated Predicate Calculus to Querying Inconsistent Databases. \ Proc. {\em Computational Logic 2000}, Springer LNCS 1861, pp.
 926-941.

 \bibitem{tplp}
Arenas,~M., Bertossi,~L. and Chomicki,~J. Answer Sets for Consistent Query Answers. \ {\em Theory and Practice of Logic Programming}, 2003, 3(4\&5):393-424.


 \bibitem{barcelo1}
Barcelo,~P. and Bertossi,~L. \
Logic Programs for Querying Inconsistent Databases. \ Proc. PADL 2003, Springer LNCS 2562, pp. 208-222.

\bibitem{bertossiSynth}
Bertossi.~L. \ {\em Database Repairing and Consistent Query Answering}. \ Synthesis Lectures in Data Management. Morgan \& Claypool, 2011.

\bibitem{tkde}
Bertossi, L. and Li, L.  \ Achieving Data Privacy through Secrecy Views and Null-Based Virtual Updates. \
{\em IEEE Trans. Knowledge and Data Engineering}, 2013,
25(5):987-1000.

\bibitem{tplp17}
Bertossi, L. and Bravo, L. \
Consistency and Trust in Peer Data Exchange Systems. \ {\em Theory and Practice of Logic Programming}, 2017, 17(2):148-204. \ Extended version as
Corr Arxiv Paper cs.DB/1606.01930.

\bibitem{tocs}
Bertossi,~L. and Salimi,~B. From Causes for Database Queries to Repairs and Model-Based Diagnosis and Back. \ {\em Theory of Computing Systems}, 2017,  61(1):191-232.

\ignore{
\bibitem{flairsExt}
Bertossi,~L. and Salimi,~B. \ Causes for Query Answers from Databases: Datalog Abduction, View-Updates, and Integrity Constraints. \ {\em Int. J. Approximate Reasoning}, 2017, 90:226-252.

\bibitem{lpnmr19}
Bertossi,~L. \
Repair-Based Degrees of Database Inconsistency. Proc. LPNMR 2019,  Springer LNCS 11481, pp. 195-209.
}

\bibitem{kais}
Bertossi,~L. \ Specifying and Computing Causes for Query Answers in Databases via Database Repairs and Repair Programs. \ {\em Knowledge and Information Systems}, 2021, 63(1):199-231.

\bibitem{deem}
Bertossi,~L., Li,~J., Schleich,~M., Suciu,~D. and Vagena,~Z. \
\newblock Causality-Based Explanation of Classification Outcomes.
\newblock In {\em Proceedings of the Fourth Workshop on Data Management for
  End-To-End Machine Learning, DEEM@SIGMOD 2020}, pages 6:1-6:10, 2020.

\bibitem{rw21}
Bertossi,~L. \ Score-Based Explanations in Data Management and Machine Learning: An Answer-Set Programming Approach to Counterfactual Analysis. \ In {\em Reasoning Web. Declarative Artificial Intelligence 2021}, Springer LNCS 13100, 2022, pp. 145-184.


\bibitem{tplp}
Bertossi,~L. \ Declarative Approaches to Counterfactual Explanations for Classification. Theory and Practice of Logic Programming. 2021. https://doi.org/10.1017/S1471068421000582. arXiv Paper 2011.07423.

\bibitem{ijclr21}Bertossi,~L. and Reyes,~G. \ Answer-Set Programs for Reasoning about Counterfactual Interventions and Responsibility Scores for Classification. \ In {\em Inductive Logic Programming}, Proc. 1st International Joint Conference on Learning and Reasoning (IJCLR'21), Springer LNAI 13191, 2022, pp. 41-56.

\bibitem{bravoJournal}
Bravo, L. and Bertossi, L. \newblock Disjunctive Deductive Databases for Computing Certain and Consistent Answers to Queries from Mediated Data Integration Systems. \newblock {\em Journal of Applied Logic}, 2005, 3(2):329-367.

\bibitem{calimeri08}
Calimeri,~F., Cozza,~S., Ianni,~G. and Leone,~N. \ Computable Functions in ASP: Theory and Implementation. Proc. ICLP 2008, Springer LNCS 5366, pp. 407-424.

 \bibitem{calimeri09}
 Calimeri,~F., Cozza,~S., Ianni,~G. and Leone,~N. \ An ASP System with Functions, Lists,and Sets. Proc. LPNMR 2009, Springer LNCS 5753, pp. 483-489.

\bibitem{monica}
Caniupan,~M. and Bertossi,~L.  \ The Consistency Extractor System: Answer Set Programs for Consistent Query Answering in Databases. \ {\em Data \& Knowledge Engineering}, 2010, 69(6):545-572.



\bibitem{CH04}
Chockler,~H. and Halpern,~J. \
\newblock Responsibility and Blame: A Structural-Model Approach.
\newblock {\em J. Artif. Intell. Res.}, 2004, 22:93-115.

\bibitem{eiterLoc}
Eiter,~T.,  Fink,~M., Greco,~G. and Lembo,~D. \
Repair Localization for Query Answering from Inconsistent Databases. \ {\em ACM Trans. Database Syst.}, 2008, 33(2):10:1-10:51.

\bibitem{zaniolo}
Giannotti,~F., Greco,~S., Sacca,~D. and Zaniolo,~C. \ Programming with Non-Determinism in Deductive Databases. \ {\em Annals of Mathematics in Artificial Intelligence}, 1997, 19(1-2):97-125.

  \bibitem{grecoLP}
Greco,~G.,  Greco,~S. and Zumpano,~E. \
A Logical Framework for Querying and Repairing Inconsistent Databases. \ {\em IEEE Trans. Knowl. Data Eng.}, 2003, 15(6):1389-1408.



\bibitem{HP05}
Halpern,~J. and Pearl,~J. \
\newblock Causes and Explanations: A Structural-Model Approach. Part I: Causes.
\newblock {\em The British Journal for the Philosophy of Science}, 2005,
  56(4):843-887.



\bibitem{leone}
Leone,~N., Pfeifer,~G., Faber,~W., Eiter,~T., Gottlob,~G., Perri,~S. and Scarcello,~F. \ The DLV System for Knowledge Representation and Reasoning. \ {\em ACM Transactions on  Computational Logic}, 2006, 7(3):499-562.


  \bibitem{suciu}
Meliou,~A., Gatterbauer,~W.,
 Moore,~K.~F. and  Suciu,~D.
\newblock The Complexity of Causality and Responsibility for Query Answers and
  Non-Answers.
\newblock Proc. VLDB 2010, pp. 34-41.



\ignore{
\bibitem{molnar}
Molnar,~C. \ {\em Interpretable Machine Learning:
A Guide for Making Black Box Models Explainable}. \ {\text https://christophm.github.io/interpretable-ml-book}, 2020.
}



\end{thebibliography}
\end{document}